# Digital Twin for Smart Societies: A Catalyst for Inclusive and Accessible Healthcare

Joshit Mohanty[a*], Sujatha Alla[b], Vaishali[c], Nagesh Bheesetty[d], Prasanthi Chidipudi[e], Satya Prakash Chowdary Nandigam[f], Marisha Jmukhadze[g], Puneeth Bheesetty[h] , Narendra Lakshmana Gowda[i]

[a,b]Engineering Management and Systems Engineering, Old Dominion University, VA, United States

[c]Biomedical Sciences, Old Dominion University, VA, United States

[d]Data Science and Information Systems, Old Dominion University, VA, United States

[e]Department of Public Health, Old Dominion University, VA, United States

[f]Department of Computer Sciences, Eastern Illinois University, United States

[g,h]Aerospace Engineering, Warsaw University of Technology, Poland

[i]Independent Researcher, United States

Email: [a]jmohanty@odu.edu, [b]salla@odu.edu, [c]vvais001@odu.edu, [d]vbhee001@odu.edu, [e]pchid001@odu.edu, [f]satyanandigam08@gmail.com, [g]m.jmukhadze@gmail.com, [h]puneeth.bheesetty@gmail.com, [i]narendra.lakshmanagowda@ieee.org

[*]Corresponding author: m.jmukhadze@gmail.com

## Abstract

With rapid digitization and digitalization, drawing a fine line between the digital and the physical world has become nearly impossible. It has become essential more than ever to integrate all spheres of life into a single Digital Thread to address pressing challenges of modern society – accessible and inclusive healthcare in terms of equality and equity. Techno-social advancements and mutual acceptance have enabled the infusion of digital models to simulate social settings with minimum resource utilization to make effective decisions. However, a significant gap exists in feeding back the models with appropriate real-time changes. In other words, active behavioral modeling of modern society is lacking, influencing community healthcare as a "whole." By creating virtual replicas of (physical) behavioral systems, digital twins can enable real-time monitoring, simulation, and optimization of urban dynamics. This paper explores the potential of digital twins to promote inclusive healthcare for evolving smart cities. We argue that digital twins can be used to:

- Identify and address disparities in access to healthcare services
- Facilitate community participation
- Simulate the impact of urban policies and interventions on different groups of people
- Aid policy-making bodies for better access to healthcare

This paper proposes several ways to use digital twins to stitch the actual and virtual societies. Several discussed concepts within this framework envision an active, integrated, and synchronized community aware of data privacy and security. The proposal also provides high-level step-wise transitions that will enable this transformation.

Keywords: Digital Twin, Smart City, Inclusivity, Healthcare Automation, Urban Policy, Artificial Intelligence.

**Introduction**

Smart cities increasingly rely on data and technology to improve urban environments' efficiency, sustainability, and livability. However, a growing concern is that smart cities can exacerbate existing inequalities and create new forms of exclusion, especially in equi-access to healthcare space (Landini et al., 2023). Digital twins are emerging as a technology that has the potential to promote inclusive access and transparency in healthcare management (Chattu et al., 2024).

The concept of the Digital Twin (DT) has been defined recently by VanDerHorn and Mahadevan as "a virtual representation of a physical system (and its associated environment and processes) that is updated through the exchange of information between the physical and virtual settings." A DT is not just a computer-generated model or data visualization of the healthcare infrastructure but a bi-directional interacting system of systems that falls under metaphysical modeling and autonomously interacts with the urban counterparts.

In the context of smart cities, digital twins can create virtual representations of urban infrastructure, such as buildings, transportation systems, energy grids, and environmental systems. These virtual representations can be used to:

- Identify and address disparities in access to urban services and opportunities
- Facilitate participatory planning and decision-making
- Simulate the impact of urban policies and interventions on different groups of people
- Develop and test innovative solutions to urban challenges

What makes DTs more intuitive is the advent and advancement of Artificial Intelligence (AI). AI enables re-visiting the existing healthcare infrastructure, population distribution, and diversity attributes to generate frameworks for policy and societal amendments, proposals, outcomes, and adversaries. Increasing efficiency in computation through low-earth satellites enables the processing of near real-time data in the urban setting (Wischert et al., 2020).

While previous studies have demonstrated the utility of digital twin technology in urban planning and healthcare services, a significant gap exists in understanding *how DT technology distinctly innovates healthcare delivery*. Current methods often focus on optimizing resource allocation or enhancing healthcare accessibility through conventional data-driven approaches. However, digital twins introduce a dynamic, *bi-directional interaction model* that facilitates real-time monitoring, predictive analytics, and scenario-based simulations tailored to diverse populations. These capabilities set digital twins apart from traditional methodologies and underscore their potential to transform healthcare services in smart cities.

**Proposal**

Digital Twin for urban healthcare planning has been on the floor for years (Thelen et al., 2022a, 2022b). The booming internet age has made it possible to count on the bits and bytes flowing across cities, helping local governments to build a digital image of the booming healthcare infra. CityZenith is leading this digital transformation through its multi-city collaboration across the U.S. Cities like New York, Phoenix, Los Angeles, and Las Vegas are a few notable examples. Orlando is a recent addition to their database to generate a complete 3-D projection of the city to create immersive public healthcare accessibility, upgrade transit systems, improve sustainable urban planning, reduce traffic congestion, and build dynamic environments for residents, among other projects. This can significantly reduce these cities' carbon footprints, increase economic viability, enhance information management (Alla et al., 2017), and provide much-needed equitable and inclusive healthcare access. Global megacities, from Singapore to Helsinki to Dubai, promptly respond to the technology and establish special administrative offices to develop digital models. Current estimations suggest a massive U.S. $280 billion in savings on urban development using the extended reality tech by 2030 (ABI Research, 2021). The island nation of Singapore has successfully demonstrated the use of the technology through its database management of millions of photos and raw data handling to recreate the city-state of about 6 million citizens virtually. It added vegetation, healthcare aid points, and underground data points to achieve the nation's sustainable development goals. Dubai, known as the City of Gold in the Middle East, has re-generated its building mappings and led to marvels like the "mirror skyscraper" and Downtown Circle, entirely planned and executed in extended reality. The digital models thus help in the virtual execution of policies, regulations, schemes, and measures that project the dynamic nature of

economics and its impact on people's lives (Ferré-Bigorra et al., 2022; Lehtola et al., 2022). The digital measures are a step toward virtual urban administration, a.k.a. the "metaverse" model for urban life. This proposal draws an analogy of this advancement to the late 1990s internet boom, where the World Wide Web failed to accommodate policies for cyber security, dark web regulation, and blockchain. The current virtual administration also fails to regulate deep internet usage headaches, resulting in illegal transactions and continuous data breaches. DTs will draw their sources from smart home systems, smart power grids, intelligent transit systems, real-time fintech systems, e-retailers, dynamic logistics, online healthcare systems, and many more. Modern-day policies and security measures are far from reality in providing a framework for effective digital administration. DTs in today's terminology are limited to 3-D models, data visualizations (Alla, 2019), machine models for data-driven optimizations (Kamuni et al., 2024, Alla et al., 2023), and approximations of human-machine interactions (Alla & Pazos 2019, Soni et al., 2024). A true DT can transform society's digital integration, promising an inclusive culture.

Digital Transformation (DT) is digitalizing urban planning, development, and maintenance processes through a slow and continuous shift from human-centric to machine-centric models. There is a paradigm shift in the way we interact with technology. Merging this transformation into our daily lives is difficult but possible. Here, the "system" is – "us," i.e., humans/society. DT of urban social aspects creates a virtual "live" environment that is termed a "Digital Twin Environment (DTE)." In this virtual environment, data flows are bidirectional with a human-centric approach. The DTE is a shared space among various stakeholders – i.e., people and organizations who have perceived interest in this virtual space. While the emergence is not considered while defining the intended contextual use, there is always a shifted interest due to political or financial situations. Thus, building a "trust" bridge between DTE and stakeholders is essential. The bridge is techno- social, techno-cultural, techno-economic, techno-psychological, etc. Eventually, this trust needs to be shifted from human-centric models to machine-centric models, which include network-centric models, model-centric models, etc. Training the machine to accommodate situational uncertainties or gut feelings in an emergent scenario is impossible. Thus, building a trust model in the DT process is essential. One facet of this argument is adapting an "ethical/moral" A.I. policy.

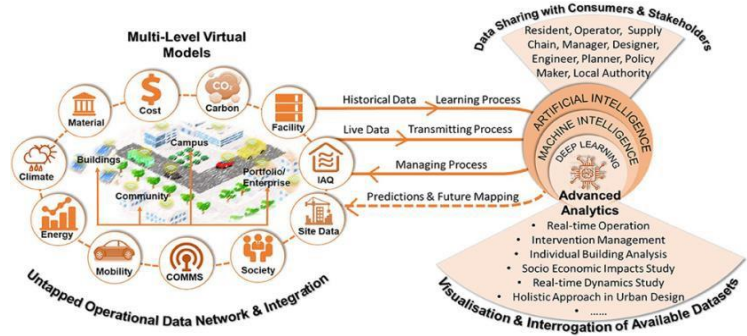

*Figure 1 | DTEs as an enabler for DTs (Created by Zhang et al., 2021).*

Moral A.I. policy seeks a transparent, open source, and privacy-centric data acquisition, sharing, transforming, selling, and targeting. A "Chief Moral A.I. Moderating Officer" must be appointed at each data processing node to regulate this. These nodes will be "check gates" between cloud and physical servers (local/edge computing). The users can range from individuals to family groups. The control level of this exchange can be set by each user (data owner). The moderating officer checks mindful data exchanges in the physical servers. Thus, even if the user accidentally relaxes their permissions, a reminder can be sent to the user based on local cluster preferences to review the permissions. This will eliminate the current one-step process where companies engage directly with the end users. This will save time for the users and generate more trust in the data-sharing policy of the companies (Ghavidel et al., 2023; Ghavidel et al., 2023). Applications range from wearable devices to smartphones, smart homes, intelligent buildings, smart societies, and even smart cities. Hence, a true DT can be achieved.

As policy-abiding bidirectional data sync is established, the domain's further bounding will be to maintain a secure trust through responsible data sourcing focusing on ethnicity, gender, culture, economics, etc. This will create an inclusive DTE and a single source of truth to verify the internet on demand. The visible internet will no longer be the cloud servers but rather every connected node to physical artifacts from which the data will be pulled. The moderating officers will form the foundation for creating the next stage – context-aware DTs in different DTEs.

We live in a growing information-sharing age where data inclusivity is "game-changing." So far, this project has developed an idea of building a virtual world for active urban planning while ensuring accessible, fair, inclusive, and privacy-centric data usage. A decentralized system is essential to maintain these attributes while mining large-scale data (Alla et al., 2017, Soltanisehat et al., 2018, Alla et al., 2024). Is "more" data relevant? Is "relevant" data freely available? The answer to both questions is no. We must, therefore, shift our focus away from data collection and evaluation principles. The previous section established that a Chief Moral A.I. Moderating Officer is key in approving data exchanges between individuals and family groups (Bonney et al., 2023). This concept can be extended to organizations or enterprises, too. Although data moderation has been discussed, the collection techniques have not been addressed today due to narrow A.I. capabilities. Current-day machine learning/A.I. can only train the machine in a way that already exists (Alla et al., 2018, July), and narrow A.I. helps roughly extrapolate it, thus relying on correlation to "suggest" the near future without any "relevancy." Hence, data relevancy needs to be addressed, like bots on the Twitter (X) platform.

This proposal moves ahead with enhancing the "relevancy" of the data collected rather than continued "mindless" mining of the data to guide urban planning. A correlation may be attained, but accuracy is absent. Hence, the cause behind the AI's suggestions remains unknown. A self-interacting platform for the infrastructure is essential to advance the free and unambiguous data flow. This can be done using X-A.I. (Explainable A.I.) to define varying weight criteria for different attributes pulled from the cloud source (Yalim et al., 2023; Mohanty et al., 2023). This is a computationally intensive task for a vast data pool. However, selecting a few characteristics that "cause" the "effect" can fix this. Initial "cause" parameters can be manually trained. Similar solutions do exist in Game Theory and Bayesian Mathematics. Quantum probability theory also offers solutions along parallel lines. All of this supports causation rather than mere correlation. Deriving causation between different factors can help establish trust, as discussed here.

To build upon this, this proposal focuses on building off of a self-moderating data-sourcing platform like Twitter (X): A Twitter (X) for the City by the Buildings. Each building is "subjectively" connected to other establishments that have a "relevancy" rather than a mere "correlation." Private property has no relevant connection to the military administrative office in the same area code but may be correlated with "societal security." Each does not depend on the other for an immediate cause; hence, no direct effect can be established. On the other hand, a hospital or shopping center tends to have more causal effects on the privately owned buildings in the neighborhood. Based on their applicability, two buildings share data through the check gate

(Dhar et al., 2022). This significantly reduces the computational load while increasing "relevancy" (Basiri et al., 2015; Abdelmagid et al., 2023).

*Digital Twin for Inclusive Healthcare*

Digital twins can be used to identify and address disparities in access to healthcare services and opportunities. For example, a digital twin of a city's emergency transportation system could be used to identify areas with limited access to public transportation. This information could then be used to plan emergency routes and dictate the type of vehicles to improve access to transportation for unserved communities.

The development of the proposed digital twin model involves several data collection and processing stages. Data acquisition includes demographic information, healthcare facility locations, transportation infrastructure, and real-time health metrics gathered through IoT-enabled devices and public health databases. This data is processed using machine learning algorithms to predict healthcare demand and optimize service distribution. Techniques such as natural language processing (NLP) are employed for sentiment analysis of public feedback, while predictive analytics models assess the impact of urban policies on healthcare equity. Advanced AI methods, including reinforcement learning, are used to simulate various healthcare scenarios and recommend actionable strategies for inclusive healthcare delivery.

Digital twins can also be used to facilitate participatory planning and decision-making. For example, a digital twin of a city's proposed development project could create a virtual environment that allows residents to visualize the project and provide feedback. This feedback could then be used to inform the design of the project.

In addition to the benefits mentioned above, DTs can also promote inclusivity and growth in smart cities in the following ways:

- Enhancing accessibility

DTs can be used to create accessible and inclusive virtual environments that people with disabilities can access (Laksminarayan et al., 2023). For example, a digital twin of a city's public transportation system could create a virtual map with information on accessible entrances, elevators, and designated seating areas. People with disabilities could use this information to plan their trips and navigate the city more easily.

The bi-directional flow of information in a DTE allows for real-time adjustments and predictive simulations based on personalized travel plans, offering a proactive rather than reactive approach to healthcare accessibility. The ability to simulate diverse urban scenarios and AI-driven insights provides unparalleled opportunities to address systemic inequities in healthcare access and delivery. Unlike traditional digital databases, where the update cycle is manual and provided in isolated settings as framed by the healthcare management, DT and DTE process information is used in conjunction with patient input data (condition), considering environmental factors, such as weather. Thus, it takes a holistic and pragmatic approach to static digital databases.

- Bridging the digital divide

DTs can be used to provide access to information and services to people who are not connected to the internet. For example, a digital twin of a city's healthcare portal could be used to create a virtual medical record library that can be accessed by people with appropriate authentication who have restricted access to healthcare at ease. This virtual library could access their medical history and other resources (Mohanty et al., 2020). Further, DTs can empower marginalized communities by giving them a voice in the city's healthcare planning and decision-making processes that affect their lives. For example, a digital twin of a city's proposed development project could create a virtual forum where residents can provide feedback. This feedback could then be used to inform the better distribution of healthcare resources.

- Enhancing sustainability

DTs can enhance the sustainability of smart cities by helping them monitor the dual impact of environmental and health. For example, a digital twin of a city's energy grid could identify and address energy inefficiencies impacting urgent care facilities. This information could then be used to implement energy-saving measures to reduce the city's carbon footprint.

**Conclusion**

DTs can transform equitable access to healthcare by creating more just and equitable urban environments. They hold potential in identifying and addressing disparities in access to urban services and opportunities by visually representing the physical world. Digital twins can also facilitate participatory planning and decision-making, simulate the impact of urban policies and interventions on different groups of people, and develop and test innovative solutions to urban challenges. As digital twins become more widely adopted, they will likely play a vital role in creating more inclusive and prosperous intelligent cities.

The proposal so far focused on decentralized data sharing in healthcare record access using technology to lay the foundations of trust in the system, a system reliable enough to design or modify future establishments accommodating empathy, diversity, and belief. Future research needs to continue to develop layers in building trust in the system to provide unbiased decision-making while ensuring the other two parameters discussed above – fairness and inclusivity. Rapid development in AI has made a promising contribution to the deployment of DTs. However, the proposal focused on some aspects of the Digital Thread for data inclusivity and fairness, and social behavior modeling is yet to be addressed.

The versatility of digital twin technology lies in its adaptability across domains, from transportation to environmental monitoring. However, its application in healthcare stands out due to the sector's intrinsic need for precision, equity, and adaptability. Digital twins provide the ability to simulate patient-centric scenarios, predict health outcomes, and design interventions tailored to marginalized communities. This capability ensures healthcare services are not only accessible but also resilient to evolving urban challenges, setting a benchmark for other domains.